\documentclass[preprint,tightenlines,floatfix,nofootinbib,superscriptaddress,fleqn]{revtex4} 
\usepackage{bm}
\usepackage{dcolumn}
\usepackage{amsmath}
\usepackage{amsfonts}
\usepackage{amssymb}
\usepackage{amscd}
\usepackage{amsthm}
\usepackage{graphicx}
\begin{document}
\preprint{PNU-NTG-01/2005}
\preprint{PNU-NuRI-01/2005}
\title{Magnetic moments of exotic pentaquark baryons}

\author{Hyun-Chul Kim}
\email{hchkim@pusan.ac.kr}
\affiliation{Department of Physics and  Nuclear Physics \& Radiation
Technology Institute (NuRI), Pusan National University, 609-735
Busan, Republic of Korea}

\author{Ghil-Seok Yang}
\email{gsyang@pusan.ac.kr}
\affiliation{Department of Physics and  Nuclear Physics \& Radiation
Technology Institute (NuRI), Pusan National University, 609-735
Busan, Republic of Korea}
\author{Micha{\l} Prasza{\l}owicz} 
\email{michal@th.if.uj.edu.pl}
\affiliation{M. Smoluchowski Institute of Physics, Jagellonian University, ul.
Reymonta 4, 30-059 Krak{\'o}w, Poland}
\author{Klaus Goeke}
\email{Klaus.Goeke@tp2.ruhr-uni-bochum.de}
\affiliation{Institut f\"ur Theoretische Physik II, Ruhr-Universit\" at Bochum,
D--44780 Bochum, Germany}
\date{January 2005}

\begin{abstract}
In this talk, we present our recent investigation on the magnetic
moments of the exotic pentaquark states, based on the chiral
quark-soliton model, all relevant intrinsic parameters being fixed by
using empirical data.
\end{abstract}
\maketitle
\section{Introduction}
Since Diakonov et al.~\cite{Diakonov:1997mm} predicted masses and
decay  widths of the exotic baryon antidecuplet within the framework
of the chiral soliton model, the findings of the exotic baryons
have been announced by many different experimental
collaborations~(see, for example, a very recent
review~\cite{Hicks:2005pm} and references therein),
though some of them are still under debate.  

Since the LEPS and CLAS collaborations used photons
to produce the $\Theta^{+}$, it is of great interest
to investigate electromagnetic properties of the exotic baryons.  In
particular, we have to know the magnetic moment of the $\Theta^{+}$
and its strong coupling constants in order to describe the 
mechanism of the pentaquark photoproduction.  However, 
information on its static properties is absent to date, so we need to
estimate them theoretically.  Recently, we calculated the magnetic
moments of the exotic pentaquarks within the framework of the chiral
quark-soliton model~\cite{Kim:2003ay,Yang:2004jr} with all relevant
dynamic parameters fixed by using empirical data.  We would like
to present those results briefly in this talk.

\section{Magnetic moments in the chiral quark-soliton model}
\label{magmoms}
The collective operator for the magnetic moments can be expressed 
in terms of six constants~\cite{Kim:1997ip,Kim:1998gt}:%
\begin{eqnarray}
\hat{\mu}^{(0)} & =& w_{1}D_{Q3}^{(8)}\;+\;w_{2}d_{pq3}D_{Qp}^{(8)}
\cdot\hat{J}_{q}\;+\;\frac{w_{3}}{\sqrt{3}}D_{Q8}^{(8)}\hat{J}_{3},\cr
\hat{\mu}^{(1)} & =& \frac{w_{4}}{\sqrt{3}}d_{pq3}D_{Qp}^{(8)}D_{8q}
^{(8)}+w_{5}\left(
  D_{Q3}^{(8)}D_{88}^{(8)}+D_{Q8}^{(8)}D_{83}^{(8)}\right)\cr
&+& w_{6}\left(
  D_{Q3}^{(8)}D_{88}^{(8)}-D_{Q8}^{(8)}D_{83}^{(8)}\right). 
\end{eqnarray}
The parameters $w_{1,2,3}$ are of order $\mathcal{O}(m_{s}^0)$,
while $w_{4,5,6}$ are of order $\mathcal{O}(m_{s})$, $m_{s}$ being
regarded as a small parameter.  Though $w_i$ can be calculated
numerically within the model, we want to fit them in this work to the 
experimental data of the octet magnetic moments.  

The full expression for the magnetic moments can be decomposed 
as three different terms:
\begin{equation}
\mu_{B}=\mu_{B}^{(0)}+\mu_{B}^{(op)}+\mu_{B}^{(wf)},
\end{equation}
where the $\mu_{B}^{(0)}$ is given by the matrix element of the
$\hat{\mu}^{(0)}$ between the purely symmetric collective states
$\left| \mathcal{R}_{J},B,J_{3}\right\rangle$~\cite{Yang:2004jr}, and
the $\mu_{B}^{(op)}$ is given as the matrix element of the
$\hat{\mu}^{(1)}$ between the symmetry states as well.  The wave
function correction $\mu_{B}^{(wf)}$ is given as a sum of the
interference matrix elements of the $\mu_{B}^{(0)}$ between purely
symmetric states and those in higher representations.  

The measurement of the $\Theta^{+}$ mass
constrains further the parameter space of the model (see
Refs.~\cite{Yang:2004jr,Goeke:2004ht} for 
details).  Denoting the set of the model parameters by
\begin{equation}
\vec{w}=(w_{1},\ldots,w_{6})
\end{equation}
the model formulae for the set of the magnetic moments in
representation $\mathcal{R}$ (of dimension $R$)
\begin{equation}
\vec{\mu}^{\mathcal{R}}=(\mu_{B_{1}},\ldots,\mu_{B_{R}})
\end{equation}
can be conveniently cast into the form of the matrix equations:%
\begin{equation}
\vec{\mu}^{\mathcal{R}}=A^{\mathcal{R}}[\Sigma_{\pi N}]\cdot\vec{w},%
\end{equation}
where rectangular matrices $A^{8}$, $A^{10}$, and $A^{\overline{10}}$
can be found in Refs.\cite{Yang:2004jr,Kim:1997ip,Kim:1998gt}.  These 
matrices depend on the pion-nucleon $\Sigma_{\pi N}$ term.    

\section{Results and discussion}\label{numres}
In order to find the set of parameters $w_{i}[\Sigma_{\pi N}]$, we
minimize the mean square deviation for the octet magnetic moments:
$\delta\mu^{8}= \sqrt{\sum_{B}\left(\mu_{B,\,{\rm th}}^{8}[\Sigma_{\pi
N}]-\mu_{B,\,{\rm exp}}^{8}\right)^{2}}/7$, where the sum extends over
all octet magnetic moments, but the $\Sigma^{0}$.  The value
$\delta\mu^{8}\simeq0.01$ is in practice independent of the
$\Sigma_{\pi N}$ in the physically interesting range 
$45\sim 75$ MeV.  Moreover, the values of the
$\mu_{B,\,th}^{8}[\Sigma_{\pi N}]$ do not depend on $\Sigma_{\pi N}$.     

Similarly, the value of the nucleon strange magnetic moment is
independent of $\Sigma_{\pi N}$ and reads $\mu_{N}^{(s)}=0.39 \,{\rm
  n.m.}$ in fair agreement with our previous analysis of
Ref.\cite{Kim:1998gt}.  Parameters $w_{i}$, however, do depend on
$\Sigma_{\pi N}$, as shown in Table.\ref{tablewi}: 
\begin{table}[h]
\begin{tabular}
[c]{|c|cccccc|}\hline
\multicolumn{1}{|c|}{$\Sigma_{\pi N}$ [MeV]} & $w_{1}$ & $w_{2}$ &
$w_{3}$ &   $w_{4}$ & $w_{5}$ & $w_{6}$ \\ \hline
$45$ & $-8.564$ & $14.983$ & $7.574$ & $-10.024$ & $-3.742$ & $-2.443$\\
$60$ & $-10.174$ & $11.764$ & $7.574$ & $-9.359$ & $-3.742$ & $-2.443$\\
$75$ & $-11.783$ & $8.545$ & $7.574$ & $-6.440$ &  $-3.742$ & $-2.443$
\\ \hline
\end{tabular}
\caption{Dependence of the parameters $w_i$ on $\Sigma_{\pi N}$.}
\label{tablewi}
\end{table}
Note that parameters $w_{2,3}$ are formally
$\mathcal{O}(1/N_{c})$ with respect to $w_{1}$. For smaller
$\Sigma_{\pi N}$, this $N_{c}$ counting is not borne by explicit
fits.  The $\mu_B^{(0)}$ can be parametrized by the following
two parameters $v$ and $w$:
\begin{equation}%
\begin{array}
[c]{ccrcc}%
v & = & \left(  2\mu_{\mathrm{n}}-\mu_{\mathrm{p}}+3\mu_{\Xi^{0}}+\mu_{\Xi
^{-}}-2\mu_{\Sigma^{-}}-3\mu_{\Sigma^{+}}\right)  /60 & = & -0.268,\\
w & = & \left(  3\mu_{\mathrm{p}}+4\mu_{\mathrm{n}}+\mu_{\Xi^{0}}-3\mu
_{\Xi^{-}}-4\mu_{\Sigma^{-}}-\mu_{\Sigma^{+}}\right)  /60 & = & 0.060.
\end{array}
\label{Eq:mean}%
\end{equation}
which are free of linear $m_{s}$ corrections~\cite{Kim:1998gt}.
This is a remarkable feature of the present fit, since when the
$m_{s}$ corrections are included, the $m_{s}$-independent parameters
need not be refitted.  This property will be used in the following
when we restore the linear dependence of the $\mu_{B}^{\overline{10}}$
on $m_{s}$.    

\begin{figure}[h]
\begin{center}
\includegraphics[scale=1.4]{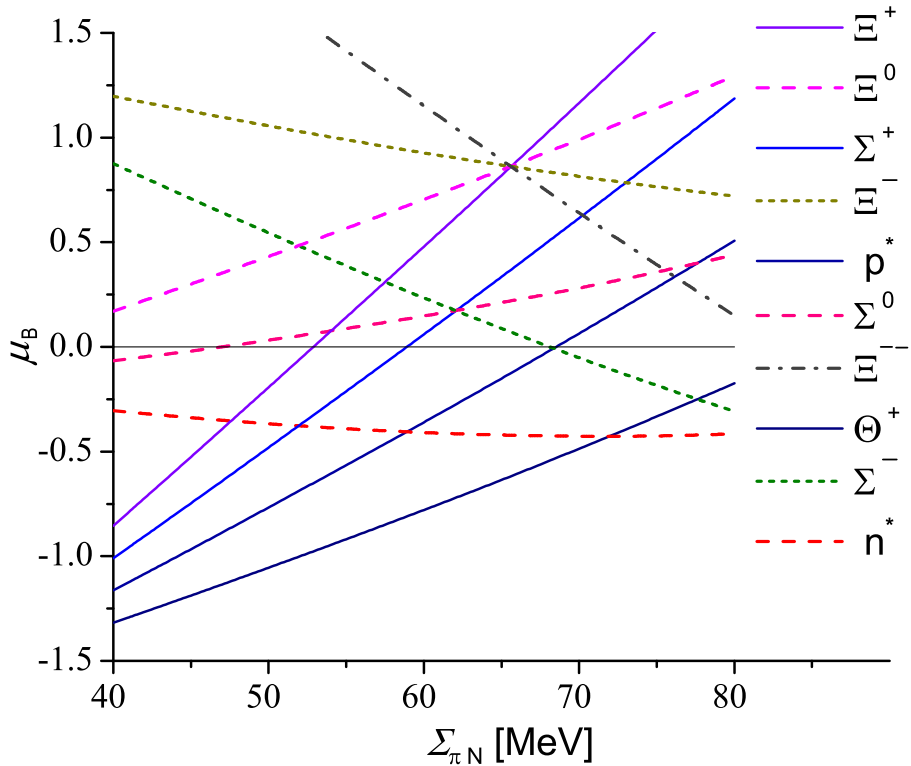}
\end{center}
\caption{Magnetic moments of antidecuplet as functions of $\Sigma_{\pi N}$.}%
\label{fig:10bar}%
\end{figure}

Finally,  for the baryon antidecuplet we have a strong dependence on 
$\Sigma_{\pi N}$, yielding the numbers listed in Table \ref{tab10b}. 
\begin{table}[h]
\begin{tabular}
[c]{|c|cccccccccc|}\hline
$\Sigma_{\pi N}$ [MeV] & $\Theta^{+}$ & $p^{\ast}$ & $n^{\ast}$
&$\Sigma_{\overline{10}}^{+}$ & $\Sigma_{\overline{10}}^{0}$ &
$\Sigma_{\overline{10}}^{-}$  & $\Xi_{\overline{10}}^{+}$ &
$\Xi_{\overline{10}}^{0}$ & $\Xi_{\overline{10}}^{-}$ &
$\Xi_{\overline {10}}^{--}$ \\ \hline 
$45$ & $-1.19$ & $-0.97$ & $-0.34$ & $-0.75$ & $-0.02$ & $\;0.71$ &
$-0.53$ & $0.30$ & $1.13$ & $1.95$ \\ 
$60$ & $-0.78$ & $-0.36$ & $-0.41$ & $\;0.06$ & $\;0.15$ & $\;0.23$ &
$\;0.48$ & $0.70$ & $0.93$ & $1.15$ \\ 
$75$ & $-0.33$ & $\;0.28$ & $-0.43$ & $\;0.90$ & $\;0.36$ & $-0.19$ &
$\;1.51$ & $1.14$ & $0.77$ & $0.39$\\ 
\hline 
\end{tabular}
\caption{Magnetic moments of the baryon antidecuplet.}
\label{tab10b}%
\end{table}
They are further depicted in Fig.\ref{fig:10bar}.
The wave function corrections cancel for the non-exotic baryons and
add constructively for the baryon antidecuplet.  In particular, for 
$\Sigma_{\pi N}=75$~MeV we have large admixture coefficient of
the 27-plet~\cite{Yang:2004jr}.  The magnetic moments of the
antidecuplet are rather small in absolute value. For
$\Theta^+$ and $p^*$ one obtains even negative values, although the
charges are positive.  As for $\Xi^{-}_{\overline{10}}$ and 
$\Xi^{--}_{\overline{10}}$ one gets positive values in spite of their 
negative signs.

\section{Conclusion and summary}\label{summary}
Our present analysis shows that $\mu_{\Theta^+}< 0$, although the
magnitude depends strongly on the model parameters.  We anticipate
that the measurement of $\mu_{\Theta^+}$ could therefore discriminate 
between different models. This also may add to reduce the ambiguities
in the pion-nucleon sigma term $\Sigma_{\pi N}$.

In the present work, we determined the magnetic moments of the
baryon antidecuplet in the \emph{model-independent} analysis
within the framework of the chiral quark-soliton model, i.e. using the
rigid-rotor quantization with the linear $m_{s}$ corrections
included.  Starting from the collective operators with dynamical
parameters fixed by experimental data, we obtained the magnetic
moments of the baryon antidecuplet.  We found that the magnetic moment
$\mu_{\Theta^{+}}$ is negative and rather strongly dependent on the
value of the $\Sigma_{\pi N}$.  Indeed, the $\mu _{\Theta^{+}}$ ranges
from $-1.19\,{\rm n.m.}$ to $-0.33\,{\rm n.m.}$ for $\Sigma_{\pi N} =
45$ and $75$ MeV, respectively.  
\section*{Acknowledgments}
HCK is grateful to the partial support for his travel by Pusan
National University.  The present work is supported by the Polish
State Committee for Scientific Research under grant 2 P03B 043 24 and
by Korean-German (KOSEF \& DFG: F01-2004-000-00102-0) and
Polish-German grants of the Deutsche Forschungsgemeinschaft. 

\end{document}